%% file: article.tex
\documentclass[%
reprint,
 amsmath,amssymb,
 aps,
 pra,
floatfix,
]{revtex4-2}

\usepackage{graphicx}
\usepackage{dcolumn}
\usepackage{bm}
\usepackage{hyperref}

\allowdisplaybreaks

\begin{document}


\title{Theory of Stationary Photon Emission from a Steadily Driven Parametric Oscillator Based on the Complex Spectral Analysis of the Heisenberg Equation}

\author{Kazuki Kanki}
\author{Satoshi Tanaka}
\affiliation{%
   Department of Physics and Nambu Yoichiro Institute of Theoretical and Experimental Physics (NITEP), Osaka Metropolitan University, Sugimoto 3-3-138, Sumiyoshi-ku, Osaka 558-8585, Japan
}%

\date{\today}

\begin{abstract}
We show how the properties of photon emission to a continuous field from a parametric oscillator relate to the behavior of the complex eigenfrequencies of the oscillator.
The parametric oscillator has complex eigenfrequencies due to non-Hermiticity with two origins: parametric amplification and dissipation resulting from the loss of photons to a continuous field.
In situations where the oscillator is close to the parametric resonance, and the coupling to the driving field is strong enough for a complex eigenfrequency to lie in the upper half of the first Riemann sheet, the number of photons of the parametric oscillator and the continuous field increases exponentially.
The parametric amplification is counteracted by dissipation due to the coupling of the parametric oscillator to the continuous field, and if the dissipation is sufficiently strong relative to the amplification, the exponential growth is suppressed.
The suppression occurs when the complex eigenfrequency responsible for the exponential growth moves into the second Riemann sheet analytically continued beyond the branch cut of the Green function on the real axis along the continuum of frequency.
As a result, there appears a stationary photon emission, where creation of photons balance with dissipation due to emission of the photons to the continuous field.
Then there is a constant flux of photons, with the number of photons in the continuous field increasing in proportion to time, while the number of photons in the parametric oscillator remains constant.

\end{abstract}

\maketitle


\section{Introduction}\label{sec:introduction}

The field exhibits quantum fluctuations in the vacuum state, with the continuous creation and annihilation of virtual particles.
Phenomena that arise from the existence of virtual photons include the Lamb shift, anomalous magnetic moments, and the Casimir force.
In the dynamical Casimir effect (DCE), a dynamical perturbation to the vacuum through time-dependent boundary conditions gives rise to amplification of the quantum fluctuations of the electromagnetic field, and virtual photons are converted into observable real photons~\cite{Moore1970,DeWitt1975,Fulling1976,Nation2012,Dodonov2020}.
The generation of real photons by moving the boundary (mirror) of an electromagnetic field requires a speed close to the speed of light, which has been difficult to achieve.
It has recently become possible to observe photons emitted by DCE using superconducting circuits to rapidly shake the effective boundary of the electromagnetic field~\cite{Wilson2011,Lahteenmaki2013,Nation2012}.

We investigate a parametric oscillator whose frequency is periodically modulated by an external field drive \cite{Law1994,Schutzhold1998}, considering the photon production by DCE as a parametric amplification of quantum fluctuations in the electromagnetic field.
The model is also applicable to parametric amplification resulting from nonlinearities in the medium, such as optical parametric oscillators \cite{Boyd2008,Walls2008,Loudon2000} and Josephson parametric oscillators~\cite{Yamamoto2008,Murch2013a,Zhong2013a,Kono2017,Aumentado2020,Blais2021,Esposito2022,Ranadive2022}.
The conversion of virtual photons into real photons is an irreversible and dissipative process, in the sense that the emitted photons never come back and are thus lost to the environment.
Therefore, we study a driven dissipative system in which a parametric oscillator is coupled with a continuous field.
We have previously obtained the complex frequencies of the discrete eigenmodes of the system, in which amplification and dissipation coexist~\cite{Tanaka2020,Tanaka2020a}.

The purpose of this paper is to theoretically clarify the non-equilibrium stationary behavior and destabilization of the system, with particular emphasis on the correspondence with the behavior of the complex engenfrequencies.
The time evolution of the system is analyzed using the Heisenberg equation for the creation and annihilation operators of all degrees of freedom.
Two causes of instability, parametric amplification and dissipation from photon loss to a continuous field, give rise to two types of non-Hermiticity in the generator (which we call the Liouvillian) of the time evolution in the Heisenberg equation.
The non-Hermiticity results in complex eigenfrequencies.
Instead of introducing a decay rate of the oscillator through phenomenological or approximate means, as is often done, e.g., in the input-output formalism~\cite{Collett1984,Gardiner1985,Clerk2010}, our treatment aims to derive the complex eigenfrequencies as the eigenvalues of the Liouvillian.
We can therefore discuss changes in the behavior of the system with regard to variations of the parameters in the Hamiltonian.
This includes parametric bifurcation in the presence of dissipation, whether the effect of dissipation is sufficiently strong to suppress exponential growth due to parametric amplification, and also the consequences of the resonance of a discrete mode frequency with a continuum.

The parametric oscillator is assumed to be constantly driven by a classical oscillating field that periodically modulates the frequency of the oscillator.
When the frequency of the driving field is nearly twice that of the oscillator, i.e., when the system is close to parametric resonance, fluctuations are amplified to the extent that the energy of the oscillator increases exponentially with time, as long as the oscillator is isolated and free from dissipation~\cite{Walls2008,Weigert2002,Diermann2019}.
This exponential growth is indicated by an eigenfrequency of the oscillator having a complex value in the upper half plane.
This amplification effect is counteracted by dissipation, which is introduced by the coupling to the continuum.
An increase in the dissipation effect reduces the magnitude of the complex eigenfrequency, which is responsible for the exponential growth.
Eventually, the complex eigenfrequency moves into the second Riemann sheet beyond the real axis, and the amplification effect is suppressed.
The need to distinguish between the first and the second Riemann sheets arises from the presence of a branch cut of the Green function along the real axis, which is a consequence of the coupling to the field with a continuum of frequency.
The suppression of exponential amplification results in stationary photon emission, in which creation of photons from parametric amplification balance with emission.

The paper is organized as follows.
In Sec.~\ref{sec:model} the model is introduced.
In Sec.~\ref{sec:Heisenberg} we give a solution of the Heisenberg equation with use of the Laplace transform.
In Sec.~\ref{sec:eigenfrequency} we discuss the behavior of eigenfrequencies.
In Sec.~\ref{sec:exp} we treat the case of instability where the parametric amplification overwhelms the dissipation.
In Sec.~\ref{sec:stationary} we discuss the behavior of the system in the regime of stationary photon emission.
We conclude the paper in Sec.~\ref{sec:conclusion}.
Some expressions and derivations are given in the Appendices.

\section{Model}\label{sec:model}

In order to describe photon creation and emission processes we consider a model
in which a parametric oscillator is coupled with continuous modes.
The Hamiltonian of the system is given by (setting $\hbar=1$ throughout)
\begin{align}\label{H(t)}
    \hat H(t)&=\omega_0\hat a^\dagger\hat a+f_0\cos\varOmega t\,(\hat a+\hat a^\dagger)^2
             \notag
             \\
             &+\int\omega\,\hat b_\omega^\dagger\hat b_\omega\,d\omega
             \notag
             \\
             &+\int g(\omega)(\hat a^\dagger\hat b_\omega+\hat b_\omega^\dagger \hat a)\,d\omega,
\end{align}
where $\hat a$ is the annihilation operator of a parametric oscillator,
and $\hat b_\omega$ is the annihilation operator of a continuous mode with frequency $\omega$.
The creation and annihilation operators satisfy the commutation relations~\cite{Loudon2000,Barnett2002},
\begin{align}
    [\hat a,\hat a^\dagger]&=1,
    \\
    [\hat b_\omega,\hat b_{\omega'}^\dagger]&=\delta(\omega-\omega').
\end{align}
The first two terms of the Hamiltonian represent the parametric oscillator
whose frequency $\omega_0$ is modulated by a periodic perturbation with frequency $\varOmega$.
We are particularly interested in the case where the system is near the parametric resonance, i.e., $\varOmega\simeq 2\omega_0$.

Let a state $|\psi(t)\rangle$ develops in time according to the Schr\"odinger equation with the Hamiltonian $\hat H(t)$:
\begin{equation}\label{SchEq}
   i\frac{d}{dt}|\psi(t)\rangle=\hat H(t)|\psi(t)\rangle.
\end{equation}
We move to the frame rotating at the frequency $\varOmega/2$ by applying the unitary transformation with
\begin{equation}
    \hat U_\mathrm{R}(t)=\exp\left[i\frac{\varOmega}{2}t\left(\hat a^\dagger\hat a+\int\!\!d\omega\,\hat b_\omega^\dagger\hat b_\omega\right)\right].
\end{equation}
so that the state becomes
\begin{equation}\label{psiR}
   |\psi_\mathrm{R}(t)\rangle=\hat U_\mathrm{R}(t)|\psi(t)\rangle.
\end{equation}
The state $|\psi_\mathrm{R}(t)\rangle$ satisfies the equation
\begin{equation}\label{SchEqRWA}
   i\frac{d}{dt}|\psi_\mathrm{R}(t)\rangle=\hat H_\mathrm{R}(t)|\psi_\mathrm{R}(t)\rangle,
\end{equation}
where the Hamiltonian $\hat H_\mathrm{R}(t)$ in the rotating frame is given by
\begin{align}\label{HR}
    \hat H_\mathrm{R}(t)&=\hat U_\mathrm{R}(t)\hat H(t)\hat U_\mathrm{R}^\dagger(t)+i\dot{\hat U}_\mathrm{R}(t)\hat U_\mathrm{R}^\dagger(t)
    \\
                     &=\hat H_\mathrm{RWA}+\hat H'(t),
    \\
    \hat H_\mathrm{RWA}&=\varDelta_0\,\hat a^\dagger\hat a+\frac{f_0}{2}(\hat a^2+{\hat a^\dagger}{}^2)
    \notag
    \\
    \label{HR2}
          &\hspace{-1em}+\int\varDelta\,\hat b_\varDelta^\dagger\hat b_\varDelta\,d\varDelta
          +\int g(\varDelta)(\hat a^\dagger\hat b_\varDelta+\hat b_\varDelta^\dagger \hat a)\,d\varDelta,
    \\
    \hat H'(t)&=\frac{f_0}{2}\left[(\hat a\hat a^\dagger+\hat a^\dagger\hat a)(e^{i\varOmega t}+e^{-i\varOmega t})\right.
    \notag
    \\
              &\hspace{8em}\left.+\hat a^2e^{-2i\varOmega t}+{\hat a^\dagger}{}^2e^{2i\varOmega t}\right],
\end{align}
where $\varDelta_0=\omega_0-\varOmega/2$ is the detuning from the parametric resonance,
and the creation and annihilation operators for the continuous modes as well as the coupling function $g(\varDelta)$
are redefined as functions of the shifted frequency $\varDelta=\omega-\varOmega/2$.
Note that the shift of each frequency by $-\varOmega/2$ is due to the second term of Eq.~\eqref{HR}:
\begin{equation}
   i\dot{\hat U}_\mathrm{R}(t)\hat U_\mathrm{R}^\dagger(t)=-\frac{\varOmega}{2}\left(\hat a^\dagger\hat a+\int\!\!d\omega\,\hat b_\omega^\dagger\hat b_\omega\right).
\end{equation}

Hereafter, we apply the rotating wave approximation by neglecting $\hat H'(t)$, each term of which oscillates rapidly in time, resulting in the time-independent Hamiltonian $\hat H_\mathrm{RWA}$, which corresponds to the following Hamiltonian in the rest frame:
\begin{align}\label{RWA1}
   \hat H_\mathrm{RWA}(t)&=\hat U_\mathrm{R}^\dagger(t)\hat H_\mathrm{RWA}\hat U_\mathrm{R}(t)
   \notag
   \\
                         &\quad\quad+\frac{\varOmega}{2}\left(\hat a^\dagger\hat a+\int\!\!d\omega\,\hat b_\omega^\dagger\hat b_\omega\right)
    \\
    &=\omega_0\hat a^\dagger\hat a+\frac{f_0}{2}(e^{i\varOmega t}\,\hat a^2+e^{-i\varOmega t}\,{\hat a^\dagger}{}^2)
    \notag
    \\
    \label{RWA}
                          &\hspace{-1em}+\int\omega\,\hat b_\omega^\dagger\hat b_\omega\,d\omega
          +\int g(\omega)(\hat a^\dagger\hat b_\omega+\hat b_\omega^\dagger \hat a)\,d\omega.
\end{align}
Thus, the Hamiltonian $H(t)$ is replaced with $H_\mathrm{RWA}(t)$ given by Eq.~\eqref{RWA} in the Schrödinger equation \eqref{SchEq} in the rest frame, and $H_\mathrm{R}(t)$ is replaced with $H_\mathrm{RWA}$ given by Eq.~\eqref{HR2} in the Schrödinger equation \eqref{SchEqRWA} in the rotating frame.

While in the rotating frame the Hamiltonian $\hat H_\mathrm{RWA}$ is independent of time and thus the energy, i.e., the expectation value of the Hamiltonian, is conserved, in the rest frame the expectation value of the Hamiltonian $\hat H_\mathrm{RWA}(t)$ depends on time as
\begin{align}\label{energy}
   \langle\psi(t)|\hat H_\mathrm{RWA}&(t)|\psi(t)\rangle=\langle\psi_\mathrm{R}(t)|\hat H_\mathrm{RWA}|\psi_\mathrm{R}(t)\rangle
   \notag
   \\
   &+\frac{\varOmega}{2}\langle\psi(t)|\hat a^\dagger\hat a+\int d\omega\,\hat b_\omega^\dagger\hat b_\omega|\psi(t)\rangle,
\end{align}
which is derived from Eq.~\eqref{RWA1}.
The first term on the right-hand side of Eq.~\eqref{energy} is time-independent because $|\psi_\mathrm{R}(t)\rangle=\exp(-i\hat H_\mathrm{RWA}t)|\psi_\mathrm{R}(0)\rangle$.
The second term, which is proportional to the expectation value of the total number of photons and is invariant under the unitary transformation \eqref{psiR}, depends on time because the total number of photons is not conserved due to the terms describing creation or annihilation of a photon pair in the Hamiltonian.
This term also shows that each photon has, on average, an energy of $\varOmega/2$.
Therefore, we can assume that a pair of photons is produced by the splitting of a single photon from the monochromatic driving field of frequency $\varOmega$.
Note that this interpretation is independent of the rotating wave approximation, since Eq.~\eqref{energy} is satisfied with $\hat H_\mathrm{RWA}(t)$ and $\hat H_\mathrm{RWA}$ replaced by $\hat H(t)$ and $\hat H_\mathrm{R}(t)$, respectively, when the approximation is not applied

In order to perform explicit calculations, we adopt a photonic crystal with nearest-neighbor transfer on the one-dimensional simple lattice as a model for the continuous field that couples to the parametric oscillator.
The Hamiltonian $H_\mathrm{PC}$ for the photonic crystal represented by the second line of Eq.~\eqref{H(t)} is given as
\begin{align}
   H_\mathrm{PC}&=\omega_\mathrm{B}\sum_{n=1}^\infty\hat b_n^\dagger\hat b_n
   -\frac{B}{2}\sum_{n=1}^\infty(\hat b_{n+1}^\dagger\hat b_n+\hat b_n^\dagger\hat b_{n+1})
   \\
                &=\int_0^\pi\omega_k\,\hat b_k^\dagger\hat b_k dk,
\end{align}
where $\hat b_n$ is the annihilation operator of a photon at the $n$-th lattice site, and $\hat b_k$ is the annihilation operator of the continuous mode with wave number $k$ ($0<k<\pi$ with the lattice constant set to be unity),
and
\begin{equation}
   \omega_k=\omega_\mathrm{B}-B\cos k.
\end{equation}
The two sets of operators $\{\hat b_n\}_{n=1,2,\dots}$ and $\{\hat b_k\}_{0<k<\pi}$ are related to each other as,
\begin{align}
   \hat b_k&=\sqrt{\frac{2}{\pi}}\sum_{n=1}^\infty\sin(kn)\hat b_n,
   \label{bk}
   \\
   \hat b_n&=\sqrt{\frac{2}{\pi}}\int_0^\pi\sin(kn)\hat b_kdk.
   \label{bn}
\end{align}
Note also that $\hat b_{\omega_k}=\hat b_k/\sqrt{B\sin k}$.
The coupling $H_\mathrm{PP}$ between the parametric oscillator and the photonic crystal represented by the third line of Eq.~\eqref{H(t)} is assumed to have the form
\begin{align}
   H_\mathrm{PP}&=g_0(\hat a^\dagger\hat b_1+\hat b_1^\dagger\hat a)
   \\
    &=g_0\int_0^\pi\sqrt{\frac{2}{\pi}}\sin k\,(\hat a^\dagger\hat b_k+\hat b_k^\dagger\hat a)\,dk,
\end{align}
where the coupling constant $g_0$ is assumed to be $g_0>0$ without losing generality.
Then, the coupling function $g(\varDelta)$ in Eq.\eqref{HR2} is given by
\begin{equation}\label{gDelta}
   g(\varDelta)=\frac{g_0}{B}\,\sqrt{\frac{2}{\pi}}\,\sqrt[4]{B^2-(\varDelta-\varDelta_\mathrm{B})^2}\,
   \theta(B-|\varDelta-\varDelta_\mathrm{B}|),
\end{equation}
where $\varDelta_\mathrm{B}=\omega_\mathrm{B}-\varOmega/2$.

\section{Solution of the Heisenberg equation via the Laplace transform}\label{sec:Heisenberg}

Hereafter we treat the time independent Hamiltonian \eqref{HR2} simply denoting it as $\hat H$.
The time evolution of the system is described in the Heisenberg picture, and an operator with a time argument denotes the corresponding one in the Heisenberg picture, e.g., $\hat a(t)=e^{i\hat Ht}\hat ae^{-i\hat Ht}$.
The Heisenberg equations for the annihilation and creation operators are given by
\begin{align}
    \label{Heisenberg1}
    i\frac{d}{dt}\hat a(t)&=\varDelta_0\,\hat a(t)+f_0\,\hat a^\dagger(t)+\int d\varDelta\,g(\varDelta)\hat b_\varDelta(t),
    \\
    \label{Heisenberg2}
    i\frac{d}{dt}\hat a^\dagger(t)&=-\varDelta_0\,\hat a^\dagger(t)-f_0\,\hat a(t)
    -\int d\varDelta\,g(\varDelta)\hat b_\varDelta^\dagger(t),
    \\
    \label{Heisenberg3}
    i\frac{d}{dt}\hat b_\varDelta(t)&=\varDelta\,\hat b_\varDelta(t)+g(\varDelta)\hat a(t),
    \\
    \label{Heisenberg4}
    i\frac{d}{dt}\hat b_\varDelta^\dagger(t)&=-\varDelta\,\hat b_\varDelta^\dagger(t)-g(\varDelta)\hat a^\dagger(t).
\end{align}

In order to solve the Heisenberg equation, we employ the Laplace transform.
We define the Laplace transform $\bar{\hat\alpha}(z)$ of a Heisenberg operator $\hat\alpha(t)$, which is any of $\hat a(t), \hat a^\dagger(t), \hat b_\varDelta(t), \hat b_\varDelta^\dagger(t)$, as
\begin{equation}
    \bar{\hat\alpha}(z)=-i\int_0^\infty e^{izt}\,\hat\alpha(t)\,dt.
\end{equation}
The Laplace transform is defined and analytic in a half-plane $\operatorname{Im}z>\sigma_0$ with some real number $\sigma_0 (\geq 0)$.
Conversely, $\hat\alpha(t)$ is given in terms of $\bar{\hat\alpha}(z)$ by the inverse Laplace transform (the Bromwich integral),
\begin{equation}\label{invLaplace}
    \hat\alpha(t)=\frac{-1}{2\pi i}\int_{-\infty+i\eta}^{+\infty+i\eta}e^{-izt}\,\bar{\hat\alpha}(z)\,dz,
\end{equation}
where $\eta>\sigma_0$.
The Laplace transform $\bar{\hat\alpha}^\dagger(z)$ of $\hat\alpha^\dagger(t)$ is related to
the Laplace transform $\bar{\hat\alpha}(z)$ of $\hat\alpha(t)$ as
\begin{equation}
    \bar{\hat\alpha}^\dagger(z)=-[\bar{\hat\alpha}(-z^*)]^\dagger.
\end{equation}

The Heisenberg equations Eq.~\eqref{Heisenberg1}--\eqref{Heisenberg4} are transformed into algebraic equations
for the Laplace transforms of the creation and annihilation operators as
\begin{align}
    \label{HeisenbergL1}
    (z-\varDelta_0)\,\bar{\hat a}(z)-f_0\,\bar{\hat a}^\dagger(z)-\int\!\!d\varDelta\,g(\varDelta)\,\bar{\hat b}_\varDelta(z)&=\hat a,
    \\
    \label{HeisenbergL2}
    (z+\varDelta_0)\,\bar{\hat a}^\dagger(z)+f_0\,\bar{\hat a}(z)+\int\!\!d\varDelta\,g(\varDelta)\,\bar{\hat b}_\varDelta^\dagger(z)&=\hat a^\dagger,
    \\
    \label{HeisenbergL3}
    (z-\varDelta)\,\bar{\hat b}_\varDelta(z)-g(\varDelta)\,\bar{\hat a}(z)&=\hat b_\varDelta,
    \\
    \label{HeisenbergL4}
    (z+\varDelta)\,\bar{\hat b}_\varDelta^\dagger(z)+g(\varDelta)\,\bar{\hat a}^\dagger(z)&=\hat b_\varDelta^\dagger.
\end{align}
Solving equations \eqref{HeisenbergL3} and \eqref{HeisenbergL4}
with respect to $\bar{\hat b}_\varDelta(z)$ and $\bar{\hat b}_\varDelta^\dagger(z)$,
we obtain
\begin{align}
    \label{bz-}
    \bar{\hat b}_\varDelta(z)&=\frac{1}{z-\varDelta}\,[g(\varDelta)\,\bar{\hat a}(z)+\hat b_\varDelta],
    \\
    \label{bz+}
    \bar{\hat b}_\varDelta^\dagger(z)&=\frac{1}{z+\varDelta}\,[-g(\varDelta)\,\bar{\hat a}^\dagger(z)+\hat b_\varDelta^\dagger].
\end{align}
Substituting Eqs.~\eqref{bz-} and \eqref{bz+} for $\bar{\hat b}_\varDelta(z)$ and $\bar{\hat b}_\varDelta^\dagger(z)$
into Eqs.~\eqref{HeisenbergL1} and \eqref{HeisenbergL2}, we obtain closed equations for $\bar{\hat a}(z)$ and $\bar{\hat a}^\dagger(z)$:
\begin{equation}
    \label{Leff}
    \left(z\mathcal I-\mathcal L_\mathrm{eff}(z)\right)
    \begin{pmatrix}
        \bar{\hat a}(z)
        \\
        \bar{\hat a}^\dagger(z)
    \end{pmatrix}
    =
    \begin{pmatrix}
        \hat a+\displaystyle\int\!\!d\varDelta\,\dfrac{g(\varDelta)}{z-\varDelta}\,\hat b_\varDelta
        \\
        \hat a^\dagger-\displaystyle\int\!\!d\varDelta\,\dfrac{g(\varDelta)}{z+\varDelta}\,\hat b_\varDelta^\dagger
    \end{pmatrix}.
\end{equation}
Here, $\mathcal I$ is the $2\times 2$ unit matrix, and the matrix $\mathcal L_\mathrm{eff}(z)$ is given by
\begin{equation}\label{Leffmatrix}
    \mathcal L_\mathrm{eff}(z)=
    \begin{pmatrix}
        \varDelta_0+\varSigma(z) & f_0
        \\
        -f_0 & -\varDelta_0-\varSigma(-z)
    \end{pmatrix},
\end{equation}
where the selfenergy $\varSigma(z)$ is defined as
\begin{equation}
    \varSigma(z)=\int\!\!d\varDelta\,\frac{g^2(\varDelta)}{z-\varDelta}.
\end{equation}
The selfenergy has a branch cut on the real axis in the interval where $z=\varDelta$ with $g(\varDelta)\neq 0$.
For the coupling function \eqref{gDelta}, the selfenergy is explicitly given as
\begin{equation}\label{selfenergy}
   \varSigma(z)=2\left(\frac{g_0}{B}\right)^2\left[z-\varDelta_\mathrm{B}\mp\sqrt{(z-\varDelta_\mathrm{B})^2-B^2}\right],
\end{equation}
where the upper (lower) sign before the square root is for $z$ in the first (second) Riemann sheet, with the two-valued square root function assigned the value such that $|z-\varDelta_\mathrm{B}-\sqrt{(z-\varDelta_\mathrm{B})^2-B^2}|<B$.

The matrix $\mathcal L_\mathrm{eff}(z)$ represents the effective Liouvillian
introduced in Refs.~\cite{Tanaka2020,Tanaka2020a}.
Multiplying Eq.~\eqref{Leff} from the left by the matrix $(z\mathcal I-\mathcal L_\mathrm{eff}(z))^{-1}$,
we obtain the solution of the Laplace transformed Heisenberg equations
for $\bar{\hat a}(z)$ and $\bar{\hat a}^\dagger(z)$ as
\begin{widetext}
\begin{equation}
    \label{az-+}
    \begin{pmatrix}
        \bar{\hat a}(z)
        \\
        \bar{\hat a}^\dagger(z)
    \end{pmatrix}
    =G(z)
    \begin{pmatrix}
        z+\varDelta_0+\varSigma(-z) & f_0
        \\
        -f_0 & z-\varDelta_0-\varSigma(z)
    \end{pmatrix}
    \begin{pmatrix}
        \hat a+\displaystyle\int\!\!d\varDelta\,\dfrac{g(\varDelta)}{z-\varDelta}\,\hat b_\varDelta
        \\
        \hat a^\dagger-\displaystyle\int\!\!d\varDelta\,\dfrac{g(\varDelta)}{z+\varDelta}\,\hat b_\varDelta^\dagger
    \end{pmatrix},
\end{equation}
\end{widetext}
where $G(z)$, which we call the Green function, is defined as
\begin{align}
    \label{Gdet}
    [G(z)]^{-1}&=\operatorname{det}[z\mathcal I-\mathcal L_\mathrm{eff}(z)]
    \\
               &\hspace{-1em}=[z-\varDelta_0-\varSigma(z)][z+\varDelta_0+\varSigma(-z)]+f_0^2,
\end{align}
and it has the properties
\begin{align}
   \label{G-}
   G(z)&=G(-z),
   \\
   \label{G*}
   [G(z)]^*&=G(z^*).
\end{align}
These properties are attributed to the symplectic property of the linear Heisenberg equation for the creation and annihilation operators~\cite{Tanaka2020,Tanaka2020a}.
The solution of the Laplace transformed Heisenberg equations
for $\bar{\hat b}_\varDelta(z)$ and $\bar{\hat b}_\varDelta^\dagger(z)$
is obtained by substituting Eq.~\eqref{az-+} for $\bar{\hat a}(z)$ and $\bar{\hat a}^\dagger(z)$
into Eqs.~\eqref{bz-} and \eqref{bz+}.

The solution of the Heisenberg equation is given by the inverse Laplace transform
from Eqs.~\eqref{bz-}, \eqref{bz+} and \eqref{az-+}, and expressed as
\begin{align}
   \label{at}
   \hat a(t)&=\alpha(t)\,\hat a+\tilde\alpha(t)\,\hat a^\dagger
    +\int\!\!d\varDelta[\beta(\varDelta;t)\,\hat b_\varDelta+\tilde\beta(\varDelta;t)\,\hat b_\varDelta^\dagger],
    \\
    \hat b_\varDelta(t)&=\hat b_\varDelta\,e^{-i\varDelta t}
    \notag
    \\
                       &\hspace{1em}+\int\!\!d\varDelta'[C(\varDelta,\varDelta';t)\,\hat b_{\varDelta'}+\tilde C(\varDelta,\varDelta';t)\,\hat b_{\varDelta'}^\dagger]
    \notag
    \\
                       &\hspace{1em}+A(\varDelta;t)\,\hat a+\tilde A(\varDelta;t)\,\hat a^\dagger,
                       \label{bt}
\end{align}
and their Hermitian conjugates.
The Laplace transforms of the coefficients in Eqs.~\eqref{at} and \eqref{bt} are given in Appendix~\ref{appendix:coefficients}.

Now we are concerned with photons created from the vacuum
and assume that the system is in the vacuum state $|0\rangle$ at the initial time $t=0$.
Here, the vacuum state $|0\rangle$ is defined by the property
\begin{equation}
   \hat a|0\rangle=0,
\end{equation}
and
\begin{equation}
   \hat b_\varDelta|0\rangle=0\ \ \text{for every }\varDelta.
\end{equation}
Then the expectation value of the number of photons in the parametric oscillator at time $t(>0)$ is
\begin{equation}
   N_a(t)\equiv\langle 0|\hat a^\dagger(t)\hat a(t)|0\rangle=|\tilde\alpha(t)|^2+\int\!\!d\varDelta|\tilde\beta(\varDelta;t)|^2,
\end{equation}
The expectation value of the number density (per unit frequency interval) of the emitted photons
is given by
\begin{equation}\label{photon_number_density}
   \langle 0|\hat b_\varDelta^\dagger(t)\hat b_\varDelta(t)|0\rangle=|\tilde A(\varDelta;t)|^2+\int\!\!d\varDelta'|\tilde C(\varDelta,\varDelta';t)|^2.
\end{equation}

\section{Complex Eigenfrequency}\label{sec:eigenfrequency}

Following Refs.~\cite{Tanaka2020,Tanaka2020a},
in this paper we refer to the generator of the time evolution of the Heisenberg equations as the Liouvillian.
Because in the model treated here the Heisenberg equation is a linear equation,
we can consider the eigenvalue problem of the Liouvillian
in a similar way as we consider the eigenvalue problem of the Hamiltonian for the Schrödinger equation.
The eigenvalues of the Liouvillian are the frequencies of the eigenmodes (eigenfrequency).
The discrete eigenvalues of the Liouvillian are the eigenvalues of the effective Liouvillian, which in our model is represented by a $2\times 2$ matrix as Eq.~\eqref{Leffmatrix}.
Note that the effective Liouvillian itself depends on the frequency, so that the eigenvalue equation is nonlinear.
The effective Liouvillian \eqref{Leffmatrix} has two types of non-Hemiticity, which give rise to complex eigenvalues.
One of these appears as the antisymmetric off-diagonal matrix elements that originate from the parametric drive.
The other is manifested in the complex values of the selfenergy, which result from the resonance of the discrete mode with the continuous field.
According to Eq.~\eqref{Gdet}, the eigenvalues of the effective Liouvillian coincide with the values of $z$
at the poles of the Green function $G(z)$.
From Eqs.~\eqref{G-} and \eqref{G*} it follows that the eigenfrequencies appear in quadruplets $(z,-z,z^*,-z^*)$,
whereas a quadruplet reduces to a pair if the eigenvalue $z$ is real and $z^*=z$,
or $z$ is pure imaginary and $z^*=-z$.

\begin{figure}[htbp]
\begin{center}
\includegraphics[width=8cm]{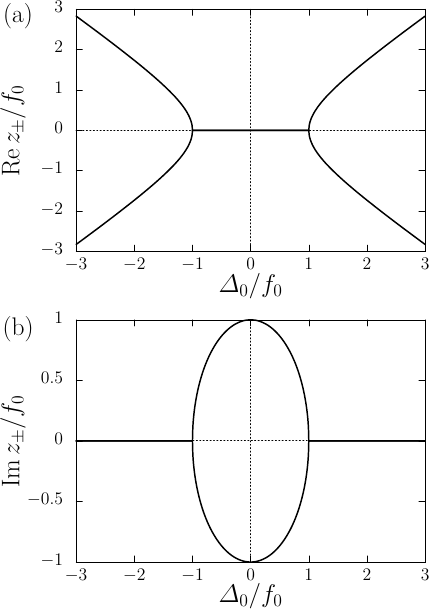}
\caption{The real (a) and imaginary (b) parts of the complex eigenfrequency of the parametric oscillator
   whose Hamiltonian is given by the first line of Eq.~\eqref{HR2}.
}
\label{Fig:paraevals}
\end{center}
\end{figure}

First, we consider the parametric oscillator decoupled from the continuum implying $g(\varDelta)\equiv 0$.
The eigenvalues of the matrix \eqref{Leff}, which is now $z$-independent because the selfenergy $\varSigma(z)$ vanishes,
give the eigenfrequencies $z_\pm=\pm\sqrt{\varDelta_0^2-f_0^2}$ shown in Fig.~\ref{Fig:paraevals}.
At the parametric bifurcation points $\varDelta_0=\pm f_0$ the real eigenvalues for $|\varDelta_0|>|f_0|$ becomes pure imaginary for $|\varDelta_0|<|f_0|$, giving rise to exponential growth of the number of photons.

\begin{figure}[htbp]
\begin{center}
\includegraphics[width=8cm]{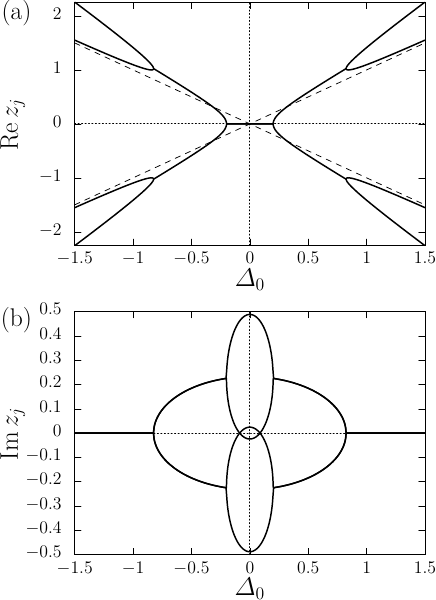}
\caption{The real (a) and imaginary (b) parts of the complex eigenfrequency $z_j$ as functions of $\varDelta_0$
   for $\varDelta_\mathrm{B}=0,\ g_0=0.3,\ f_0=0.2$ in units in which $B=1$.
   The dashed lines in (a) represent $\operatorname{Re}z_j=\pm\varDelta_0$,
   which the eigenfrequencies in the first Riemann sheet approach as $|\varDelta_0|\to\infty$.
}
\label{Fig:evals2}
\end{center}
\end{figure}

\begin{figure}[htbp]
\begin{center}
\includegraphics[width=8cm]{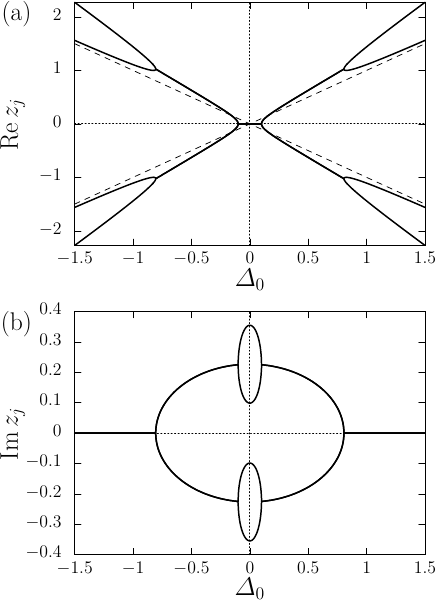}
\caption{The real (a) and imaginary (b) parts of the complex eigenfrequency $z_j$ as functions of $\varDelta_0$
   for $\varDelta_\mathrm{B}=0,\ g_0=0.3,\ f_0=0.1$ in units in which $B=1$.
   The dashed lines in (a) represent $\operatorname{Re}z_j=\pm\varDelta_0$,
   which the eigenfrequencies in the first Riemann sheet approach as $|\varDelta_0|\to\infty$.
}
\label{Fig:evals1}
\end{center}
\end{figure}

Next, for the case where the parametric oscillator is coupled to the continuum and the selfenergy is given by Eq.~\eqref{selfenergy}, we show in Fig.~\ref{Fig:evals2} and \ref{Fig:evals1} the real and imaginary parts of the complex eigenfrequencies as a function of the detuning $\varDelta_0$.
The coupling to the driving field represented by the parameter $f_0$ is larger in Fig.~\ref{Fig:evals2} than in Fig.~\ref{Fig:evals1}, with other parameters the same.

There are four eigenfrequencies at each $\varDelta_0$.
When $|\varDelta_0|>B$ and thus the frequency of the parametric oscillator is outside the continuum band, all the eigenfrequencies are real.
Two of the real eigenfrequencies are in the first Riemann sheet, and as $|\varDelta_0|\to\infty$ asymptotically approach the asymptotes $z_j=\pm\varDelta_0$ represented by the dashed lines in the Figs.~\ref{Fig:evals2}(a) and \ref{Fig:evals1}(a).
As $\varDelta_0$ decreases,
each real eigenfrequency in the first Riemann sheet reaches the band edge of the continuous mode frequency,
then it goes into the second Riemann sheet, and collides with another eigenfrequency.
At this point, the two eigenvectors as well as the two eigenfrequencies of each pair coalesce and thus form an exceptional point.
For $|\varDelta_0|$ less than the value at the exceptional point, each coalesced pair of eigenvectors bifurcates into a pair of eigenvectors with complex conjugate eigenvalues.
Thus, the eigenfrequencies change from real to complex due to the resonance of the discrete frequency of the parametric oscillator with the continuum of frequencies of the continuous modes in an infinite space.
The imaginary part of an eigenvalue gives the decay rate of the corresponding component of the parametric oscillator.

On further decreasing $|\varDelta_0|$ and $|\varDelta_0|\approx f_0$, there appear another kind of exceptional points,
where the parametric bifurcation occurs.
Since the parametric bifurcation point is in the second Riemann sheet, parametric instability, i.e., exponential amplification, does not occur near the bifurcation point.
If the coupling $f_0$ to the driving external field is strong enough, the complex eigenvalues eventually may cross the real axis at smaller $|\varDelta_0|$ and enter the first Riemann sheet (e.g. Fig.~\ref{Fig:evals2}), giving rise to instability with exponential amplification.
Otherwise, if none of the complex eigenvalues does not reach the real axis after the parametric bifurcation (e.g. Fig.~\ref{Fig:evals1}), the system remains stable, meaning there is no exponential growth of physical quantities.

\section{Photon Emission}

Whether the system is stable or unstable is indicated by the absence or presence of complex frequencies in the first Riemann sheet.
In this section we examine the photon emission in each case.

\subsection{Exponential Growth of the Number of Photons in the Unstable Region}\label{sec:exp}

An eigenfrequency in the upper half plane of the first Riemann sheet gives rise to exponential growth of the number of photons.
This makes the system unstable.
The exponentially growing component of the number density of emitted photons \eqref{photon_number_density} comes from the residue at the pole $z=i\gamma_\mathrm{I}$ ($\gamma_\mathrm{I}>0$) in the first Riemann sheet of the Green function $G(z)$ in Eqs.~\eqref{tildeAt} and \eqref{tildeCt}.
We define the spectrum of the exponentially growing component of the emitted photons as
\begin{equation}\label{Fexp}
   F_\mathrm{exp}(\varDelta)=\lim_{t\to\infty}\langle 0|\hat b_\varDelta^\dagger(t)\hat b_\varDelta(t)|0\rangle
   e^{-2\gamma_\mathrm{I}t},
\end{equation}
which converges because the number density of photons diverges proportionally to $e^{2\gamma_\mathrm{I}t}$.
It follows from Eqs.~\eqref{photon_number_density}, \eqref{tildeAt} and \eqref{tildeCt} that
\begin{equation}\label{Lorentzian}
   F_\mathrm{exp}(\varDelta)\propto\frac{g^2(\varDelta)}{\varDelta^2+\gamma_\mathrm{I}^2}.
\end{equation}

\begin{figure}[htbp]
\begin{center}
\includegraphics[width=8cm]{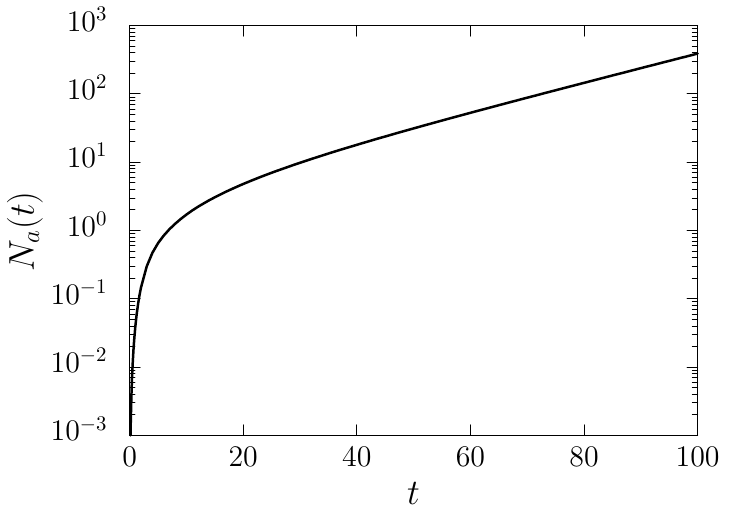}
\caption{
   The number of photons in the parametric oscillator as a function of time $t$
   for $\varDelta_0=0,\ \varDelta_\mathrm{B}=0,\ g_0=0.3,\ f_0=0.2$ in units in which $B=1$.
}
\label{Fig:nphoton2}
\end{center}
\end{figure}

\begin{figure}[htbp]
\begin{center}
\includegraphics[width=8cm]{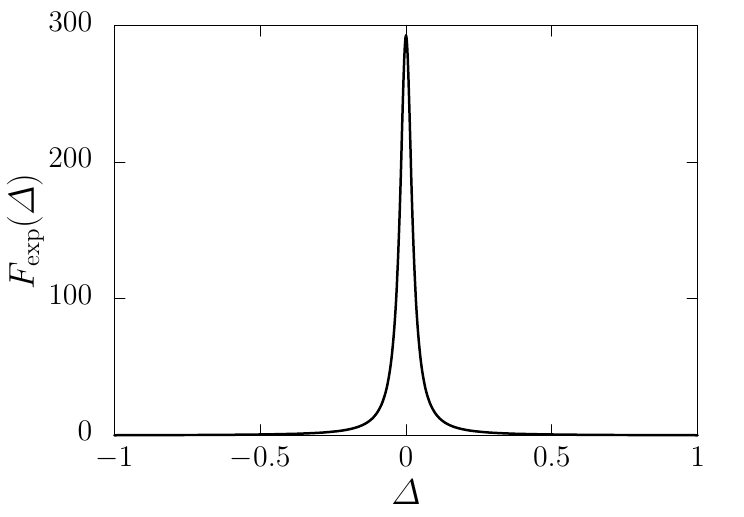}
\caption{
   The photon emission spectrum $F_\mathrm{exp}(\varDelta)$ defined by Eq.~\eqref{Fexp}
   for $\varDelta_0=0,\ \varDelta_\mathrm{B}=0,\ g_0=0.3,\ f_0=0.2$ in units in which $B=1$.
}
\label{Fig:Fexp}
\end{center}
\end{figure}

In Fig.~\ref{Fig:nphoton2} we show a result for the time-evolution of the number of photons in the parametric oscillator.
The number of photons increases proportionally to $e^{2\gamma_\mathrm{I}t}$ after an initial time region.
In Fig.~\ref{Fig:Fexp} is shown the corresponding emitted photon spectrum defined by Eq.~\eqref{Fexp}.
The function has the form of Eq.~\eqref{Lorentzian} and the width of the peak is given by $\gamma_\mathrm{I}$.

In practice, the divergence in the unstable region is eventually suppressed by nonlinearity and other effects not included in the model considered in this paper.
However, we will not discuss further possible behaviors in the unstable region when factors not included in the model are considered.
In any case, as described in the next subsection, away from the parametric resonance, the complex frequency in the first Riemann sheet responsible for the exponential growth goes into the second Riemann sheet, and then the system becomes stable.

\subsection{Stationary Photon Emission\\ in the Stable Region}\label{sec:stationary}

If the system is away from the parametric resonance at $\varDelta_0=0$
and/or the coupling to the external driving field is small enough,
the parametric amplification effect is suppressed and there is no exponential growth.
In this case there is no complex eigenfrequency in the first Riemann sheet.
In the long-time region the number of photons of the parametric oscillator converges to a constant, implying that generation and emission of photons are balanced.
Then, there is a constant flux of photons, and the number of emitted photons increases linearly with time.

\begin{figure}[htbp]
\begin{center}
\includegraphics[width=8cm]{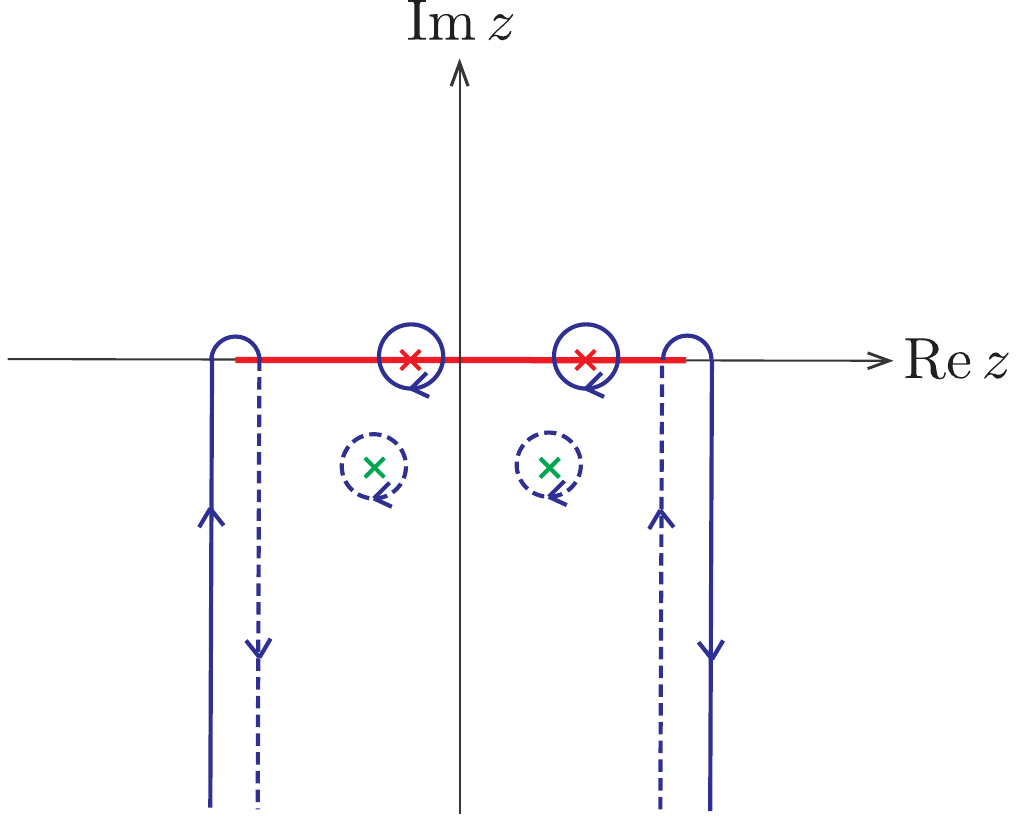}
\caption{
   Contours of the integrals of the inverse Laplace transform with integrands given by Eqs.~\eqref{tildealphat}--\eqref{Ct}.
   This figure is for the case $\omega_\mathrm{B}=0$, and there is a branch cut of the Green function, indicated by the thick line, in the interval $-B<\operatorname{Re}z<B$ on the real axis.
   There are three kinds of contributions to the integrals:
   The contributions from the poles in the lower half of the second Riemann sheet, indicated by the dashed circles.
   The contributions from the poles on the real axis in the sheet analytically continued from the upper half of the first Riemann sheet, indicated by the solid circles.
   The contours encircling the photonic band edges give the branch point effect.
   In these contours, the solid lines are in the first Riemann sheet and the dashed lines are in the second Riemann sheet.
}
\label{Fig:srs}
\end{center}
\end{figure}

For the integrals of the inverse Laplace transforms with integrands given by Eqs.~\eqref{tildealphat}--\eqref{Ct}, the integration contours are deformed as depicted in Fig.~\ref{Fig:srs}.
Note that the Green function has two branch cuts on the real axis associated with $\varSigma(z)$ and $\varSigma(-z)$ in Eq.~\eqref{Gdet}, and that the two cuts coincide when $\omega_\mathrm{B}=0$, which is the case in the figure.
There are three kinds of contributions to the integrals.
The contributions from the poles of the Green function in the lower half plane of the second Riemann sheet decrease exponentially to zero.
The integrals along the contours encircling the photonic band edges, which give the branch point effect, decrease to zero according to power laws.
In the long-time limit, there remain the contributions from the poles of the integrands on the real axis.
For examples, $\tilde\alpha(t)\to 0$ as $t\to\infty$ because its Laplace transform given by Eq.~\eqref{tildealphat} has no pole on the real axis, for $\tilde\beta(\varDelta;t)$ with the Laplace transform given by Eq.~\eqref{tildebetat} there remains a contribution from the pole at $z=-\varDelta$, and for $\tilde C(\varDelta,\varDelta';t)$ with the Laplace transform given by Eq.~\eqref{tildeCt} there remain contributions from the two poles at $z=\varDelta,-\varDelta'$.

The constant number of photons in the parametric oscillator in the long-time limit is obtained by using the asymptotic form \eqref{aasymp} as
\begin{equation}\label{photon_number}
    \lim_{t\to\infty}\langle 0|\hat a^\dagger(t)\hat a(t)|0\rangle=
    f_0^2\int\!\!d\varDelta\,g^2(\varDelta)\left|G(\varDelta+i0^+)\right|^2,
\end{equation}
which comes from the contribution to the integral for the inverse Laplace transform from the pole at $z=-\varDelta$ of Eq.~\eqref{tildebetat}.
Note that $G(\varDelta+i0^+)$ means the value of $G(z)$ at $z=\varDelta$ in the Riemann sheet analytically continued from the upper half-plane.
As shown in Appendix \ref{appendix:emission} the number density of photons increases linearly with time in the long-time region, and we define the photon emission spectrum $F_{b^\dagger b}(\varDelta)$ as
\begin{equation}\label{emission_spectrum}
   F_{b^\dagger b}(\varDelta)=\lim_{t\to\infty}\frac{\langle 0|\hat b_\varDelta^\dagger(t)\hat b_\varDelta(t)|0\rangle}{t}.
\end{equation}
As derived in Appendix~\ref{appendix:emission}, we obtained the following expression for this function:
\begin{equation}\label{Fnd}
   F_{b^\dagger b}(\varDelta)=2\pi f_0^2g^2(\varDelta)g^2(-\varDelta)\left|G(\varDelta+i0^+)\right|^2.
\end{equation}
Note that in cases for large values of $|\varDelta_0|$ where there are real discrete eigenfrequencies (in the first Riemann sheet), the photon number density $\langle 0|\hat b_\varDelta^\dagger(t)\hat b_\varDelta(t)|0\rangle$ has also an oscillating component.

The function $F_{b^\dagger b}(\varDelta)$ is an even function of $\varDelta$, which can be seen with Eqs.~\eqref{G-} and \eqref{G*}.
This is because photons are emitted in pairs each of which consists of two photons with frequencies $\varDelta_+,\varDelta_-$ that satisfy $\varDelta_++\varDelta_-=0$ in the rotating frame.
In the rest frame the frequencies $\omega_+,\omega_-$ of such a pair add up to the frequency of the external driving field $\varOmega$: $\omega_++\omega_-=\varOmega$.
In other words, each pair of photons is assumed to be produced by the splitting of a single photon from the monochromatic pumping field.
The process is called parametric down-conversion.

The emitted photons propagate toward the spatial infinity.
The photon flux is given by the expectation value of the photon flux operator,
\begin{align}
   &\hat J_{n\to n+1}(t)
   \notag
   \\
   &=i\frac{B}{2}\left[\{\hat b_{n+1}^\to(t)\}^\dagger\,\hat b_n^\to(t) - \{\hat b_n^\to(t)\}^\dagger\,\hat b_{n+1}^\to(t)\right],
\end{align}
which corresponds to the number of photons going from the $n$-th site to the $(n+1)$-th site per unit time.
The superscript $\to$ to the creation and annihilation operators means that the right moving components are taken as \cite{Clerk2010}
\begin{equation}
   \hat b_n^\to=-\frac{i}{\sqrt{2\pi}}\int_0^\pi dk\,e^{ikn}\,\hat b_k,
\end{equation}
from $\hat b_n$ given by Eq.~\eqref{bn}.
In this way it is assumed that we take the limit of infinity in the system size before taking the long-time limit, so that the outgoing waves never reach the far end of the system and reflect back.
We obtain the constant part of the photon flux in the long-time limit, which is independent of $n$ as well as $t$, as
\begin{equation}
   J_\to=2\pi f_0^2\int d\varDelta\,g^2(\varDelta)g^2(-\varDelta)\left|G(\varDelta+i0^+)\right|^2,
\end{equation}
by calculating $\lim_{t\to\infty}\langle 0|\hat J_{n\to n+1}(t)|0\rangle$ with use of Eq.~\eqref{ndexp} retaining its singular term (containing a delta-function), which lead to the behavior that in the long time region the photon number density per unit interval of frequency increases in proportion to time, as discussed in Appendix~\ref{appendix:emission}.
Hence, with Eqs. \eqref{emission_spectrum} and \eqref{Fnd} we can see that in the long-time region the constant part of the photon flux is equal to the rate of increase of the total number of emitted photons.

\begin{figure}[htbp]
\begin{center}
\includegraphics[width=8cm]{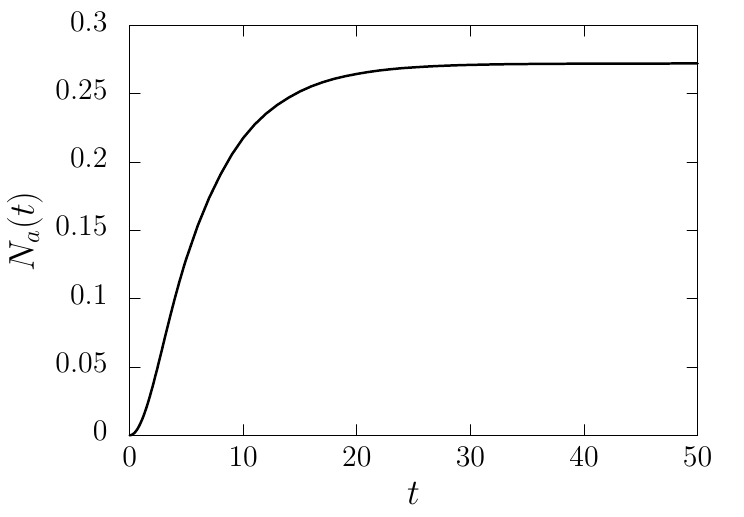}
\caption{
   The number of photons in the parametric oscillator as a function of time $t$
   for $\varDelta_0=0,\ \varDelta_\mathrm{B}=0,\ g_0=0.3,\ f_0=0.1$ in units in which $B=1$.
}
\label{Fig:nphoton1}
\end{center}
\end{figure}

\begin{figure}[htbp]
\begin{center}
\includegraphics[width=8cm]{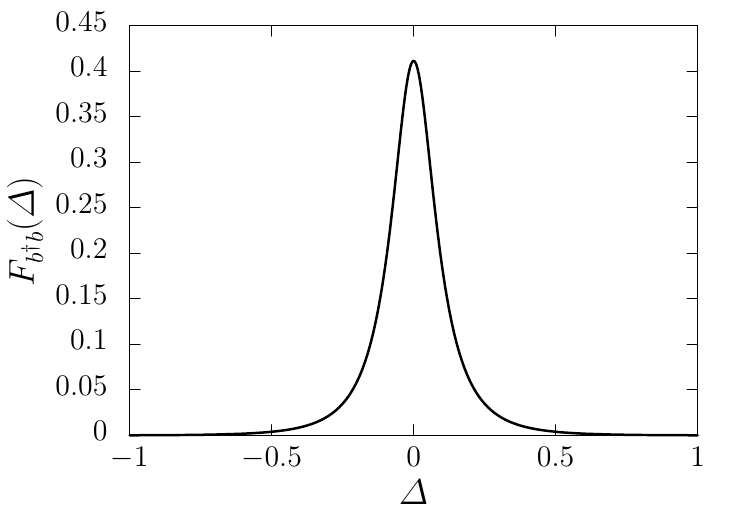}
\caption{
   The photon emission spectrum $F_{b^\dagger b}(\varDelta)$ given by Eq.~\eqref{Fnd}
   for $\varDelta_0=0,\ \varDelta_\mathrm{B}=0,\ g_0=0.3,\ f_0=0.1$ in units in which $B=1$.
}
\label{Fig:Fnd1}
\end{center}
\end{figure}

Now we show the behavior of the system in the case of Fig.~\ref{Fig:evals1}, where the coupling to the driving field is reduced from that in Fig.~\ref{Fig:evals2}, so that the coupling to the continuous field is strong enough to suppress the exponential growth and bring the parametric oscillator to a steady state even when the parametric resonance condition of the parametric oscillator is satisfied ($\varDelta_0=0$).
In Fig.~\ref{Fig:nphoton1} we show a result for the time evolution of the number of photons in the parametric oscillator, which converges to a constant in the long time limit.
In Fig.~\ref{Fig:Fnd1} is shown the corresponding emitted photon spectrum given by Eq.~\eqref{Fnd}.

\begin{figure}[htbp]
\begin{center}
\includegraphics[width=8cm]{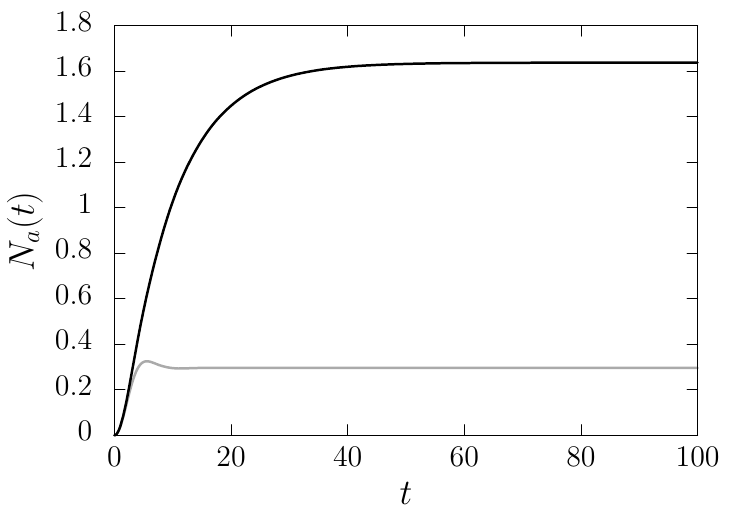}
\caption{
   The number of photons in the parametric oscillator as a function of time $t$
   for $\varDelta_0=0.15$ (black line) and $\varDelta=0.3$ (gray line) in units in which $B=1$.
   The other parameters are $\varDelta_\mathrm{B}=0,\ g_0=0.3,\ f_0=0.2$.
}
\label{Fig:npo}
\end{center}
\end{figure}

\begin{figure}[htbp]
\begin{center}
\includegraphics[width=8cm]{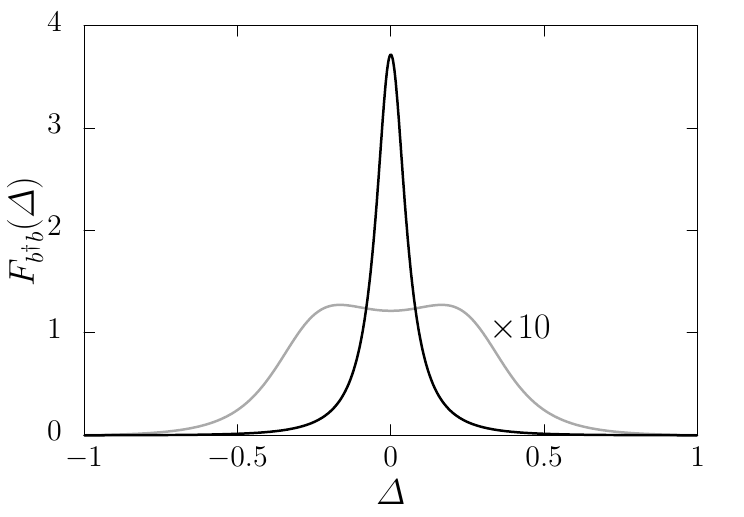}
\caption{
   The photon emission spectrum $F_{b^\dagger b}(\varDelta)$ given by Eq.~\eqref{Fnd}
   for $\varDelta_0=0.15$ (black line) and $\varDelta_0=0.3$ (gray line, multiplied by 10) in units in which $B=1$.
   The other parameters are $\varDelta_\mathrm{B}=0,\ g_0=0.3,\ f_0=0.2$.
}
\label{Fig:Fnd2}
\end{center}
\end{figure}

Even in cases where the system is unstable at the parametric resonance ($\varDelta_0=0$), stationary photon emission occurs in the long time region if the system is far from the parametric resonance and has not reached the threshold for the exponential amplification.
The threshold $\varDelta_{0\mathrm{c}}$ in terms of the detuning $\varDelta_0$ is such that there are complex eigenfrequencies in the first Riemann sheet when $|\varDelta_0|<\varDelta_{0\mathrm{c}}$, and is given by $\varDelta_{0\mathrm{c}}=\sqrt{f_0^2-4g_0^4/B^2}$.
In the case of Fig.~\ref{Fig:evals2}, $\varDelta_{0\mathrm{c}}\approx 0.0872$ in units of $B$.
In this case, for two values of $\varDelta_0 (>\varDelta_{0\mathrm{c}})$, the time evolution of the number of photons in the parametric oscillator is shown in Fig.~\ref{Fig:npo}, and the photon emission spectrum $F_{b^\dagger b}(\varDelta)$ is shown in Fig.~\ref{Fig:Fnd2}.
For $\varDelta_0=0.15$ the real parts of the eigenfrequencies in the lower half plane of the second Riemann sheets
are both zero, and the photon emission spectrum has a single peak at $\varDelta=0$.
The degenerate value of the real parts of the eigenfrequencies splits at the parametric bifurcation points, and the photon emission spectrum is double-peaked for $\varDelta_0=0.3$.
This correspondence between the bifurcation of the complex eigenfrequencies and the splitting of a peak in the photon emission spectrum can be understood in the following way.
The photon emission spectrum, given by Eq.~\eqref{Fnd}, is proportional to the absolute squared value of the Green function.
Therefore, the complex eigenfrequencies, which correspond to the poles of the Green function, determine the positions of the peaks in the photon emission spectrum.
It should also be remarked that the intensity of the photon emission is significantly reduced for the larger value of $\Delta_0$ in the far region from the parametric resonance prior to the parametric bifurcation.
These properties of the photon emission spectrum near the parametric bifurcation points appear to be well reflected in the ``sparrow-tail'' structure observed in an experiment of the dynamical Casimir effect~\cite{Lahteenmaki2013}.

\section{Conclusion}\label{sec:conclusion}

We studied the photon emission to a continuous field from a parametric oscillator.
The parametric oscillator is assumed to be driven by a monochromatic classical field with a frequency close to twice the oscillator frequency.
We applied a rotating wave approximation for coupling to the driving field.
Then, in the rotating frame, the Hamiltonian is time-independent, and complex eigenfrequencies are obtained as values at the poles of the Green function for the Heisenberg equation.

When the parametric amplification dominates the dissipation, the numbers of photons in the parametric oscillator and the continuous field increase exponentially with time.
The exponential growth is attributed to a complex eigenfrequency lying in the first Riemann sheet.
As the effect of dissipation increases relative to the effect of parametric amplification, the complex eigenfrequencies in the first Riemann sheet approach the real axis and eventually cross the real axis to enter the second Riemann sheet.
In situations where there is no complex eigenfrequency in the first Riemann sheet, the exponential growth of fluctuations is suppressed, and in the long-time region the number of photons in the parametric oscillator is constant as a consequence of the balance between parametric amplification and photon emission.
Then the system is in a stationary photon emission regime, where the number of photons emitted to the continuous field increases in proportion to time, and there is a constant photon flux.

In the parametric amplification process, photons are created and annihilated in pairs.
A pair of generated photons as an excitation of the parametric oscillator is converted into a pair of photons of the continuous field.
Frequencies of a pair of emitted photons add up to the frequency of the driving field.
In other words, a pump photon with the frequency of the driving field is down-converted into a pair of photons.
Thus, the photon emission spectrum is symmetric around half the frequency of the driving field.

In contrast to the case of a parametric oscillator free from dissipation, which exhibits parametric bifurcation points that separate stable and unstable regions, when a parametric oscillator is coupled with a continuous field and thus dissipation is introduced, a parametric bifurcation occurs in the second Riemann sheet and the system is stable around the parametric bifurcation point.
The parametric bifurcation is accompanied by a splitting of the peak in the photon emission spectrum, which is expected to give rise to a ``sparrow-tail'' structure observed experimentally~\cite{Lahteenmaki2013}.

Beyond the parametric bifurcation point and in the vicinity of the parametric resonance, if a complex eigenfrequency in the lower half of the second Riemann sheet crosses the real axis and enters the first Riemann sheet, the system becomes unstable, resulting in an exponential growth of the number of photons in the parametric oscillator and the continuous field.
Therefore, the threshold to instability of the driven dissipative system studied in this paper is where a complex eigenfrequency crosses the real axis in the complex plane.

A photon-pair creation operator in the Hamiltonian leads to formation of coherent pair-superposition states, which are quadrature-squeezed states~\cite{Loudon2000}.
In these states, the continuous-mode field is entangled.
We hope to discuss squeezing and entanglement in detail in future publications.
We will also discuss the approach to stationary photon emission and the way the system becomes unstable when a complex eigenfrequency in the second Riemann sheet in the lower-half complex plane reaches the real axis.

\begin{acknowledgments}
   This work was supported by JSPS KAKENHI Grant Numbers JP18K03496, JP24K06901.
\end{acknowledgments}

\appendix

\section{The coefficients in the solution of the Heisenberg equation}\label{appendix:coefficients}

The coefficients in Eqs.~\eqref{at} and \eqref{bt} are obtained by the inverse Laplace transform, defined as Eq.~\eqref{invLaplace}, from their respective Laplace transforms, which are denoted with a bar and given as follows:
\begin{align}
   \bar{\tilde\alpha}(z)&=f_0 G(z),
   \label{tildealphat}
    \\
   \bar{\tilde\beta}(\varDelta;z)&=-f_0 G(z) \frac{g(\varDelta)}{z+\varDelta},
    \label{tildebetat}
    \\
   \bar\alpha(z)&=G(z)[z+\varDelta_0+\varSigma(-z)],
   \label{alphat}
    \\
    \label{betat}
   \bar\beta(\varDelta;z)&=
   G(z)[z+\varDelta_0+\varSigma(-z)]\frac{g(\varDelta)}{z-\varDelta},
   \\
   \bar{\tilde A}(\varDelta;z)&=f_0 G(z) \frac{g(\varDelta)}{z-\varDelta},
   \label{tildeAt}
    \\
   \bar{\tilde C}(\varDelta,\varDelta';z)&=-f_0 G(z)\frac{g(\varDelta)}{z-\varDelta}\frac{g(\varDelta')}{z+\varDelta'},
    \label{tildeCt}
    \\
    \label{At}
   \bar A(\varDelta;z)&=
    G(z)\left[z+\varDelta_0+\varSigma(-z)\right]\frac{g(\varDelta)}{z-\varDelta},
    \\
    \label{Ct}
   \bar C(\varDelta,\varDelta';z)&=
    G(z)\left[z+\varDelta_0+\varSigma(-z)\right]\frac{g(\varDelta)}{z-\varDelta}\frac{g(\varDelta')}{z-\varDelta'}.
\end{align}

\section{The long-time asymptotic forms of the Heisenberg operators $\hat a(t)$ and $\hat b_\varDelta(t)$}\label{appendix:asymp}

The asymptotic forms of the photon annihilation operators $\hat a(t)$ and $\hat b_\varDelta(t)$ in the long-time limit are given by
\begin{widetext}
\begin{align}
   \hat a(t)&\simeq\int\!\!d\varDelta\,g(\varDelta)G(\varDelta+i0^+)[\varDelta+\varDelta_0+\varSigma(-\varDelta-i0^+)]e^{-i\varDelta\,t}\,\hat b_\varDelta
            -f_0\int\!\!d\varDelta\,g(\varDelta)G(-\varDelta+i0^+)e^{i\varDelta\,t}\,\hat b_\varDelta^\dagger,
   \label{aasymp}
   \\
   \hat b_\varDelta(t)&\simeq e^{-i\varDelta\, t}\left[G(\varDelta+i0^+)\left\{\left(\varDelta-\varDelta_0-\varSigma(\varDelta-i0^+)\right)\left(\varDelta+\varDelta_0+\varSigma(-\varDelta-i0^+)\right)+f_0^2\right\}\hat b_\varDelta
   \right.
   \notag
   \\
                      &\hspace{3em}+2\pi if_0g(\varDelta)g(-\varDelta)G(\varDelta+i0^+)\,\hat b_{-\varDelta}^\dagger
                      \notag
                      \\
                      &\hspace{3em}+g(\varDelta)G(\varDelta+i0^+)[\varDelta+\varDelta_0+\varSigma(-\varDelta-i0^+)]\,\hat a
                       +f_0g(\varDelta)G(\varDelta+i0^+)\,\hat a^\dagger\Bigr]
                      \notag
                      \\
                      &+\int\!\!d\varDelta'\left[C_\mathrm{ns}(\varDelta,\varDelta';t)\,\hat b_{\varDelta'}+\tilde C_\mathrm{ns}(\varDelta,\varDelta';t)\,\hat b_{\varDelta'}^\dagger\right]
                      \label{basymp},
\end{align}
\end{widetext}
where $C_\mathrm{ns}(\varDelta,\varDelta';t)$ and $\tilde C_\mathrm{ns}(\varDelta,\varDelta';t)$
are non-singular functions of $\varDelta'$, i.e., they do not contain a delta-function.
These formulas are derived as follows, based on the fact that in the long time limit only the poles on the real axis contribute to the Bromwich integrals of the inverse Laplace transform from the Laplace transforms of the coefficients given by Eqs.~\eqref{tildealphat}--\eqref{Ct}.

In Eq.~\eqref{at} for $\hat a(t)$, $\alpha(t)$ and $\tilde\alpha(t)$ vanish in the long-time limit,
because the integrands of the Bromwich integrals for them do not have any pole on the real axis.
By retaining the residues at $z=\varDelta$ in Eq.~\eqref{betat} and at $z=-\varDelta$ in Eq.~\eqref{tildebetat},
we obtain Eq.~\eqref{aasymp}.

In Eq.~\eqref{bt} for $\hat b_\varDelta(t)$, the singular terms of the integrands over $\varDelta'$ are separated using Eq.~\eqref{sinc} as
\begin{align}
   &\tilde C(\varDelta,\varDelta';t)
   \notag
   \\
   &\simeq
   -f_0\frac{g(\varDelta)g(\varDelta')}{\varDelta+\varDelta'}
   \notag
   \\
   &\hspace{1em}\times\left[e^{-i\varDelta\,t}G(\varDelta+i0^+)-e^{i\varDelta't}G(-\varDelta'+i0^+)\right]
   \notag
   \\
   &=f_0g(\varDelta)g(\varDelta')G(\varDelta+i0^+)
   \notag
   \\
   &\hspace{2em}\times e^{i(-\varDelta+\varDelta')t/2}
   \frac{2i\sin\left(\dfrac{\varDelta+\varDelta'}{2}t\right)}{\varDelta+\varDelta'}
   \notag
   \\
   &\hspace{1em}+(\text{non-singular terms})
   \notag
   \\
   &\simeq 2\pi if_0g(\varDelta)g(-\varDelta)G(\varDelta+i0^+)e^{-i\varDelta\,t}\delta(\varDelta+\varDelta')
   \notag
   \\
   &\hspace{1em}+(\text{non-singular terms}),
   \label{tCasymp}
\end{align}
and
\begin{align}
   &C(\varDelta,\varDelta';t)
   \notag
   \\
   &\simeq
   \frac{g(\varDelta)g(\varDelta')}{\varDelta-\varDelta'}
   \notag
   \\
   &\hspace{1em}\times\bigl[e^{-i\varDelta\,t}G(\varDelta+i0^+)\{\varDelta+\varDelta_0+\varSigma(-\varDelta-i0^+)\}
   \notag
   \\
   &\hspace{1em}-e^{-i\varDelta't}G(\varDelta'+i0^+)\{\varDelta'+\varDelta_0+\varSigma(-\varDelta'-i0^+)\}\bigr]
   \notag
   \\
   &=g(\varDelta')g(\varDelta)G(\varDelta+i0^+)\{\varDelta+\varDelta_0+\varSigma(-\varDelta-i0^+)\}
   \notag
   \\
   &\hspace{1em}\times e^{-i(\varDelta+\varDelta')t/2}\frac{-2i\sin\left(\dfrac{\varDelta-\varDelta'}{2}t\right)}{\varDelta-\varDelta'}
   \notag
   \\
   &\hspace{1em}+(\text{non-singular terms})
   \notag
   \\
   &\simeq -2\pi ig^2(\varDelta)G(\varDelta+i0^+)\{\varDelta+\varDelta_0+\varSigma(-\varDelta-i0^+)\}
   \notag
   \\
   &\hspace{1em}\times e^{-i\varDelta\,t}\delta(\varDelta-\varDelta')
   \notag
   \\
   &\hspace{1em}+(\text{non-singular terms}),
   \label{Casymp}
\end{align}
where uses have been made of the foramula
\begin{align}\label{sinc}
   \lim_{t\to\infty}\frac{\sin\left(\dfrac{\omega}{2}t\right)}{\omega}
   &=\lim_{t\to\infty}\frac{1}{2}\int_{-t/2}^{t/2}e^{i\omega t'}dt'
   \notag
   \\
   &=\pi\delta(\omega).
\end{align}
Substituting Eqs.~\eqref{tCasymp} and \eqref{Casymp} into Eq.~\eqref{bt},
and retaining the residues of the poles at $z=\varDelta$ of the integrands for $A(\varDelta;t)$ and $\tilde A(\varDelta;t)$,
we obtain Eq.~\eqref{basymp}.

\section{Derivation of Eq.~\eqref{Fnd} for the photon emission spectrum $F_{b^\dagger b}(\varDelta)$}\label{appendix:emission}
In the expression \eqref{photon_number_density} for the number density of the emitted photons,
$\tilde A(\varDelta;t)$ converges as $t\to\infty$ to a constant
given by the contribution to the integral for the inverse Laplace transform from the pole at $z=\varDelta$ of Eq.~\eqref{tildeAt}.
For $\tilde C(\varDelta,\varDelta';t)$ in the long time limit there remain contributions to the integral for the inverse Laplace transform from the two poles at $z=\varDelta,-\varDelta'$ of Eq.~\eqref{tildeCt} as
\begin{align}
   &\tilde C(\varDelta,\varDelta';t)\simeq
   \notag
   \\
   &-f_0 g(\varDelta)g(\varDelta')\frac{G(\varDelta+i0^+)e^{-i\varDelta\,t}-G(-\varDelta'+i0^+)e^{i\varDelta't}}{\varDelta+\varDelta'}
   \notag
   \\
   &=-f_0 g(\varDelta)g(\varDelta')G(\varDelta+i0^+)\frac{e^{-i\varDelta\,t}-e^{i\varDelta't}}{\varDelta+\varDelta'}
   \label{C1}
   \\
   &-f_0 g(\varDelta)g(\varDelta')e^{i\varDelta't}\frac{G(\varDelta+i0^+)-G(-\varDelta'+i0^+)}{\varDelta+\varDelta'}
   \label{C2}.
\end{align}
The term \eqref{C1}, which is proportional to
\begin{equation}\label{C1t}
   \frac{e^{-i\varDelta\,t}-e^{i\varDelta't}}{\varDelta+\varDelta'}
   =-2ie^{-i(\varDelta-\varDelta')t/2}\frac{\sin\left(\dfrac{\varDelta+\varDelta'}{2}t\right)}{\varDelta+\varDelta'},
\end{equation}
is a singular function of $\varDelta+\varDelta'$ in the limit $t\to\infty$ in the sense of Eq.~\eqref{sinc}.
In order to obtain the number density of photons,
we substitute \eqref{C1}$+$\eqref{C2} into the integrand in Eq.~\eqref{photon_number_density}.
The squared absolute value of the term \eqref{C2} is time-independent.
The product of \eqref{C1} and the complex conjugate of \eqref{C2} have the factor \eqref{C1t},
and the integral of it remain finite as $t\to\infty$, as can be seen with Eq.~\eqref{sinc}.
Integrating over $\varDelta'$ the squared absolute value of \eqref{C1} gives the $t$-linear term
of the number density of photons \eqref{photon_number_density} in the long time limit, as
\begin{align}
    &f_0^2 g^2(\varDelta) |G(\varDelta+i0^+)|^2 \int\!\!d\varDelta' g^2(\varDelta')
    \left|\frac{e^{-i\varDelta t}-e^{i\varDelta't}}{\varDelta+\varDelta'}\right|^2
    \notag
    \\
    =&f_0^2 g^2(\varDelta) |G(\varDelta+i0^+)|^2 \int\!\!d\varDelta' g^2(\varDelta')
    \frac{4\sin^2\!\!\left(\dfrac{\varDelta+\varDelta'}{2}t\right)}{(\varDelta+\varDelta')^2}
    \notag
    \\
    \approx&2\pi t f_0^2 g^2(\varDelta) g^2(-\varDelta) |G(\varDelta+i0^+)|^2,
    \label{tFnd}
\end{align}
where use has been made of the formula \cite{Cohen-Tannoudji1998}
\begin{equation}\label{tdelta}
   \frac{\sin^2\!\!\left(\dfrac{\omega}{2}t\right)}{\omega^2}\approx\frac{\pi}{2}t\,\delta(\omega)\quad(t\to\infty),
\end{equation}
which follows from Eq.~\eqref{sinc}.

The function $F_{b^\dagger b}(\varDelta)$ can also be obtained by using the following formula:
\begin{align}
   \lim_{t\to\infty}\langle 0|\hat b_\varDelta^\dagger(t)\hat b_{\varDelta'}(t)|0\rangle
   &=2\pi F_{b^\dagger b}(\varDelta)\delta(\varDelta-\varDelta')
   \notag
   \\
   &\hspace{1em}+(\text{non-singular terms}).
   \label{ndexp}
\end{align}
The fact that Eq.~\eqref{ndexp} gives the same definition of $F_{b^\dagger b}(\varDelta)$ as Eq.~\eqref{emission_spectrum} can be seen by letting $\varDelta'$ be equal to $\varDelta$ and using
\begin{equation}
   2\pi\delta(\varDelta-\varDelta'=0)\approx\left.\int_{-t/2}^{t/2}e^{i(\varDelta-\varDelta')t'}dt'\right|_{\varDelta=\varDelta'}=t,
\end{equation}
for large but finite $t$, which follows from Eq.~\eqref{sinc}.
Calculating the expectation value in Eq.~\eqref{ndexp} with the asymptotic form \eqref{basymp} of the continuous-mode operator, we obtain Eq.~\eqref{Fnd}.

\input{article.bbl}

\end{document}

%% file: article.bbl
%